\documentclass[letterpaper,twocolumn,superscriptaddress,floatfix]{revtex4}
\usepackage[latin1]{inputenc}
\usepackage{bm}
\usepackage{multirow,amssymb,amsbsy,amsmath}
\usepackage{graphicx}
\usepackage{verbatim}
\makeatletter
\usepackage{pifont}
\makeatother

\begin{document}

\title{Experimental detection of quantum coherent evolution through the violation of Leggett-Garg-type inequalities}

\author{Zong-Quan Zhou}
\affiliation{Key Laboratory of Quantum Information, University of Science and Technology of China,
CAS, Hefei, 230026, China}
\affiliation{Synergetic Innovation Center of Quantum Information and Quantum Physics, University of Science and Technology of China, Hefei, 230026, China}
\author{Susana F. Huelga$\footnote{email:susana.huelga@uni-ulm.de}$}
\affiliation{Institut f\"{u}r Theoretische Physik, Albert-Einstein-Allee 11, Universit\"{a}t Ulm,
D-89069 Ulm, Germany}
\author{Chuan-Feng Li$\footnote{email:cfli@ustc.edu.cn}$}
\affiliation{Key Laboratory of Quantum Information, University of Science and Technology of China,
CAS, Hefei, 230026, China}
\affiliation{Synergetic Innovation Center of Quantum Information and Quantum Physics, University of Science and Technology of China, Hefei, 230026, China}
\author{Guang-Can Guo}
\affiliation{Key Laboratory of Quantum Information, University of Science and Technology of China,
CAS, Hefei, 230026, China}
\affiliation{Synergetic Innovation Center of Quantum Information and Quantum Physics, University of Science and Technology of China, Hefei, 230026, China}
\date{\today}

\begin{abstract}
{We discuss the use of inequalities of the Leggett-Garg type (LGtI) to witness quantum coherence and present the first experimental violation of this type of inequalities using a light-matter interfaced system. By separately benchmarking the Markovian character of the evolution and the translational invariance of the conditional probabilities, the observed violation of a LGtI is attributed to the quantum coherent character of the process. These results provide a general method to benchmark `quantumness' when the absence of memory effects can be independently certified and confirm the persistence of quantum coherent features within systems of increasing complexity.}
\end{abstract}


\maketitle

The formulation of quantitative coherence measures has received renewed attention following the observation of coherent behavior in rather complex systems \cite{fleming}.
Recent theoretical approaches \cite{tilman,florian} aim at developing a rigorous framework for the quantification of coherence and have provided computable measures in terms of certain functionals of the density matrix. Here we discuss a different approach where quantum coherent evolution is assessed in terms of the violation of inequalities of the Leggett-Garg type (LGtI). Inequalities of this form have their origin in the celebrated approach by Leggett-Garg \cite{macro,lgi,lgi3}, as discussed below, and have been suggested as a useful tool to discriminate classical versus quantum transport \cite{Lambert2010,nori,wilde} and to witness non-Markovianity \cite{mauro}. Here we discuss how LGtI can be used to benchmark coherent quantum behavior when supplemented by additional tests. Our approach allows the experimental refutation of a large class of classical Markovian theories while confirming the validity of the quantum mechanical predictions.

In its original formulation \cite{macro,lgi,lgi3}, Leggett-Garg inequalities aimed at providing a quantitative criteria to characterize the boundary between the quantum and classical domains by testing the persistence of coherent effects on a macroscopic scale. In practice, the main experimental challenge for this type of tests comes from the implementation of truly noninvasive measurements (NIM), which are assumed to be possible if a system is truly macroscopic. Experimental tests involving weak measurements \cite{superc,photon2,spin2} can, in principle, minimize the `quantum mechanically invasiveness' \cite{clive1}. Other schemes involving ideal negative measurements have recently allowed for the experimental tests on the Leggett-Garg inequality in microscopic systems \cite{weak1,weak2}.
Independent of the question of coherence at the macroscopic level, the Leggett-Garg approach also lies at the heart of discussions on the similarity and difference between spatial and temporal correlations in quantum mechanics \cite{Brukner2004,Marcovitch2011} and the ability to perform quantum computation \cite{Brukner2004,Morikoshi2006}.

In our work, we depart from the original formulation of Leggett-Garg, and consider instead LGtI that are derived from different premises and will therefore have a different scope. That is, our aim is to witness coherent behavior without specific reference to macroscopically distinguishable states.
To avoid the requirement of performing noninvasive measurements at intermediate times, we adopt the approach presented in Refs. \cite{s1,s2,spin1} where testable inequalities bounding the system's autocorrelation functions are obtained on the basis of the assumption of {\em stationarity}. According to this assumption, the conditional probability $Q_{ij}(t_1,t_2)$ to find a system in state $j$ at time $t_2$, if it is in state $i$ at time $t_1$ only depends on the time difference $(t_2-t_1)$. When stationarity holds, LG-like inequalities can then be formulated as follows \cite{s1}:
\begin{equation}
\mathbf{K_{\mp}}=K(0,2t)\mp 2K(0,t)\geq-1,\\
\label{tbi1}
\end{equation}
where $K(t_1,t_2)$ denotes the autocorrelation function defined as $K(t_1,t_2)=\langle M(t_1)M(t_2)\rangle$ for a certain dichotomic observable $M(t)$. Inequalities Eqs.(\ref{tbi1}) are violated by quantum mechanical unitary dynamics and are easily testable by projective measurements.

While the assumption of stationarity immediately leads to easily testable inequalities, it was known to narrow down the class of macrorealist theories which are put to the test \cite{s1} and it is only recently that its implications have become fully understood \cite{clive1}. Namely, provided that the system can be initialized in a well-defined state, Eqs.(\ref{tbi1}), which have the same functional form as those resulting from imposing NIM, hold on the basis of time-translational invariance of the probabilities and Markovianity of the system evolution \cite{clive1,spin1}. Remarkably, these are assumptions that can be tested independently, as opposed, for instance, to the situation encountered with standard Bell inequalities, where the so called {\em fair sampling assumption} is not directly testable \cite{fair}. As a result, if stationarity does hold for the considered experimental setup, we argue that the inequalities of the form Eqs.(\ref{tbi1}) provide a quantitative way to witness the persistence of coherent effects, that is, they allow for benchmarking `quantumness' \cite{gisin}.

Here we present the first experimental violation of this form of LGtI with complete tests of the stationary assumption by generating and probing a delocalized atomic excitation over two macroscopically separated crystals. The dynamical evolution of the atomic excitation is controlled with a polarization-dependent atomic-frequency-comb (AFC) \cite{AFC08,AFC09,polarization,ref1,ref2,g2,entangle} and will be monitored through the violation of LGtI. The validity of the stationary assumption in this system is experimentally verified with independent tests to assess both the time translational invariance of the conditional probabilities and the Markovianity of the evolution.

The experimental sample, which is similar to our memory hardware presented in Ref. \cite{polarization}, is composed of two pieces of Nd$^{3+}$:YVO$_4$ crystals (5-ppm doping level, 3-mm thickness) sandwiching a $45^\circ$ half-wave plate (HWP). Horizontally ($H$)- and vertically ($V$)-polarized photons with wavelength of approximately 880 nm can be independently processed by the first and second crystal, respectively \cite{polarization,SHB}. Spectroscopic properties of the sample are provided in the Supplemental Material \cite{SI}.

\begin{figure*}[t]
\begin{center}
\includegraphics [width= 5 in]{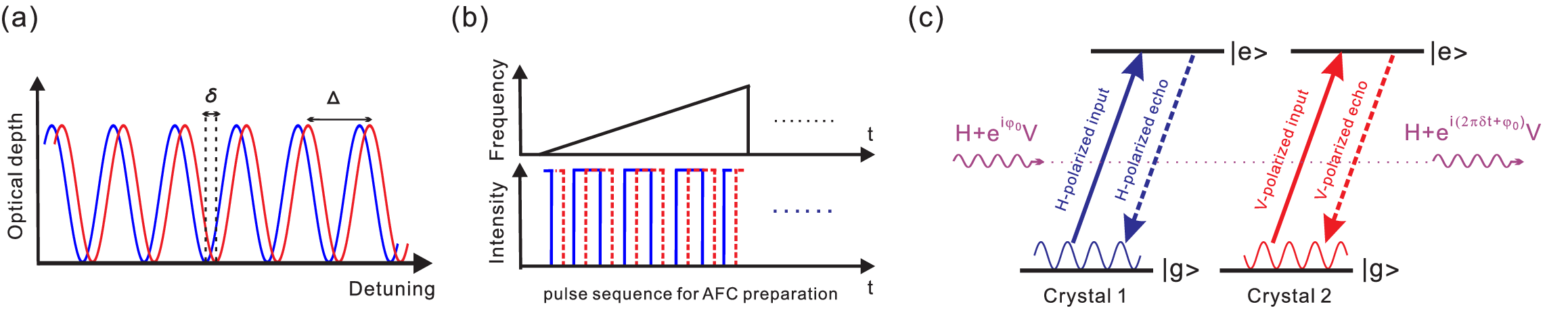}
\end{center}
\caption{(Color online) (a). The $H$- (blue) and $V$- (red) polarized AFC are prepared with a periodicity of $\Delta$ and a relative detuning of $\delta$. (b). The pulse sequence for AFC preparation used in the experiment. The frequency of pump light is scanned in each cycle while the intensities of $H$- (blue solid line) and $V$- (red dashed line) polarized light are modulated periodically. (c). The input photons with polarization states of $H+e^{i\varphi_0}V$ are resonant with the $^{4}I_{9/2}\rightarrow{ }^4F_{3/2}$ transition of Nd$^{3+}$. The $H-$ and $V-$ polarized components are absorbed and stored as atomic excitations in the first and second crystal, respectively. After the evolution in the two crystals, the readout photons will have rotated polarization states. The comb structure is prepared by frequency selective optical pumping of atoms from the ground state $|g \rangle$ to an auxiliary Zeeman state (not shown).}

\end{figure*}

\begin{figure*}[t]
\begin{center}
\includegraphics [width= 5 in]{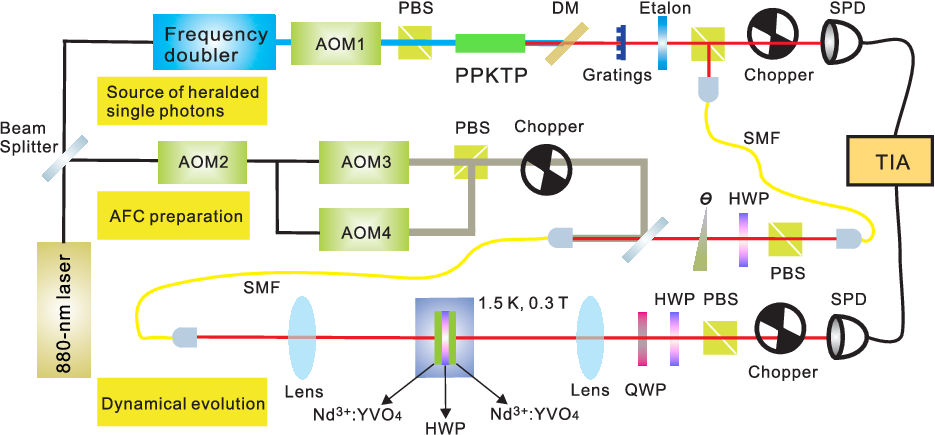}
\end{center}
\caption{(Color online) Experimental setup for violation of LGtI with delocalized single-photon excitation in two macroscopically separated crystals. The photon pairs are generated in the PPKTP crystal. The pump source is obtained by sending the master laser to a frequency doubler. The blue pump light is modulated into pulses by the acousto-optic modulators (AOM1) and then removed by the dichroic mirror (DM). The photon pairs are spectrally filtered by the gratings and the etalon. The two photons in pair are separated by the polarization beam splitter (PBS). The $H$-polarized photons are directed to a single-mode fiber (SMF)-coupled single-photon detector (SPD). The AOM3 and AOM4 generate the $H$- and $V$-polarized pump light for $H$- and $V$-polarized AFC preparation, respectively. The signal photons' polarization is controlled by the PBS, the half-wave plate (HWP) and the phase plate ($\theta$). The AFC preparation light and single photons are combined with a beam splitter and collected into a SMF. After a programmable delay, the polarization of retrieved signal photons is analyzed with a quarter-wave plate (QWP), a HWP and a PBS. The single photon signals are analyzed with time interval analyzers (TIA). Mechanical choppers are used to protect the SPD from classical light. Further details are provided in Ref. \cite{SI}.}
\end{figure*}

The AFC protocol requires a tailored absorption profile with a series of periodic absorbing peaks separated by $\Delta$ (see Fig. 1). We first consider that a single photon interacts with a single AFC. Note that it is also possible to characterize the two-level AFC using a linear electric susceptibility. However, we choose to describe the medium using a collective Dicke-state conditioned on the absorption of a photon \cite{AFC09}, since it is a straightforward and easy way to understand the experiment. The atomic state with and without the photon excitation can be represented by $|e\rangle_N=\sum_j^Nc_{j}e^{-ikz_j} e^{i2\pi\delta_jt}|g_1\cdot\cdot\cdot e_j \cdot\cdot\cdot g_N\rangle$ \cite{AFC09} and $|g\rangle_N=|g_1\cdot\cdot\cdot g_j \cdot\cdot\cdot g_N\rangle$, respectively. Here $N$ is the total number of atoms in the comb; $|g_j\rangle$ ($|e_j\rangle$) represents the ground (excited) state of atom $j$; $z_j$ is the position of atom $j$; $k$ is the wave number of the input field; $\delta_j$ is the detuning of the atom with respect to the light frequency and the amplitude $c_j$ depends on the frequency and on the position of atom $j$. If a single photon is absorbed at $t=0$, the collective atomic excitation will rapidly dephase because each item in $|e\rangle_N$ acquires a phase $ e^{i2\pi\delta_jt}$ depending on the specific detuning of each atom. However, due to the periodical structure of AFC, $\delta_j\simeq m_j \Delta$ with $m_j$ integer. A strong rephasing echo occurs after a time $\tau_s=1/\Delta$ \cite{AFC09}.

With the two AFC of the two crystals being prepared with the same periodicity $\Delta$, when the input photon's polarization is chosen as $H+e^{i\varphi_0}V$, the AFC of the two crystals will be excited simultaneously. As shown in Fig. 1(c), the $H-$ and $V-$polarized components are independently stored and acquire a different phase shift in the first and second crystals, respectively. The collective excitation in the two crystals can be represented by a coherent superposition of the form $|\Psi(t)\rangle=1/\sqrt{2}(|e\rangle_{N1}|g\rangle_{N2}+|g\rangle_{N1}|e\rangle_{N2} e^{i(2\pi\delta t+\varphi_0)})$ where $N1(N2)$ is the number of involved atoms in the first (second) crystal. $\delta$ is the frequency detuning between the two AFC.

We define the dichotomic observable $M(t)=|D\rangle\langle D|-|A\rangle\langle A|$, where the linear superposition state
$|D\rangle=1/\sqrt{2}(|e\rangle_{N1}|g\rangle_{N2}+|g\rangle_{N1}|e\rangle_{N2})$ defines the basis state with eigenvalue equal to +1 and the orthogonal state $|A\rangle=1/\sqrt{2}(|e\rangle_{N1}|g\rangle_{N2}-|g\rangle_{N1}|e\rangle_{N2})$ corresponds to the basis state with eigenvalue equal to -1. The atomic state in two crystals can now be written in a compact form as,
 \begin{eqnarray}\label{eq1}
|\Psi(t)\rangle &= \cos(\varphi/2)|D\rangle-i \sin(\varphi/2)|A\rangle,
\end{eqnarray}
where $\varphi=2\pi\delta t+\varphi_0$. To probe the atomic state, the atomic excitation is converted back to single photon through the AFC echo emission. The atomic state $|D\rangle$ and $|A\rangle$ corresponds to photonic polarization states of $H+V$ and $H-V$, respectively.

Fig. 2 shows the experimental setup for investigating the evolution of collective atomic excitation. Photon pairs are produced with a type-II spontaneous parametric down conversion (SPDC) process in a periodically poled potassium titanyl phosphate (PPKTP) crystal. The nonlinear crystal is pumped by pulsed light at 440 nm, yielding degenerate photon pairs with the $H$-polarized idler ($i$) photons and the $V$-polarized signal ($s$) photons. The pulse sequence for AFC preparation is shown in Fig. 1(b), in which the frequency of pump light is swept over 100 MHz in a 500 $\mu$s cycle and its amplitude has been modulated periodically. The $H$- and $V$-polarized lights are independently programmed to give the desired polarization-dependent AFC structures.

\begin{figure}[tbph]
\begin{center}
\includegraphics [width= 1 \columnwidth]{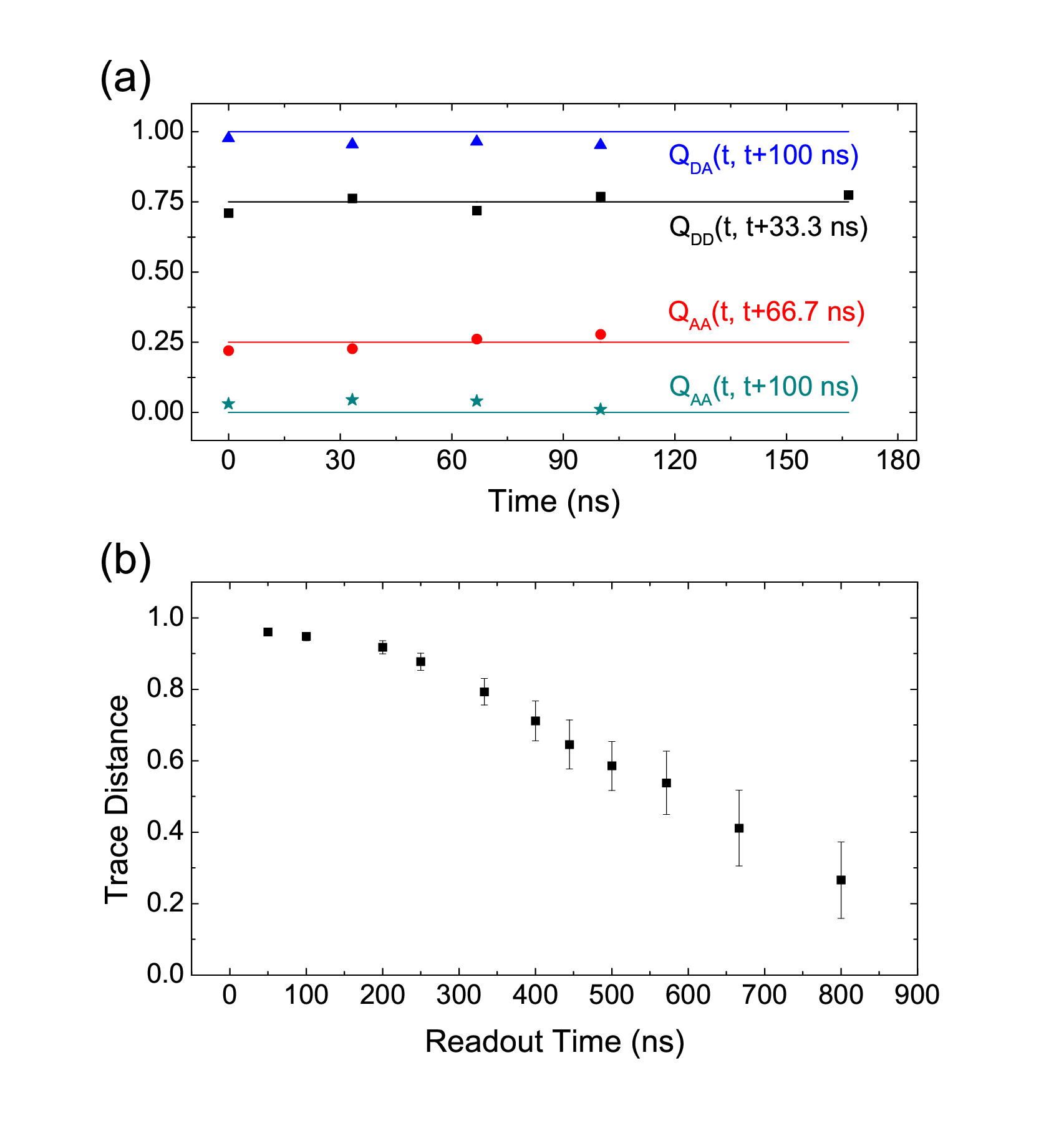}
\end{center}
\caption{(Color online) (a). Experimental test on the validity of the stationary assumption. The conditional probabilities $Q_{DA}(t,t+100$ ns) (blue triangles), $Q_{DD}(t,t+33.3$ ns) (black squares), $Q_{AA}(t,t+66.7$ ns) (red dos) and $Q_{AA}(t,t+100$ ns) (green stars) show little dependence on the evolution time $t$. Solid lines are the theoretical predictions. (b). Experimental determination of the trace distance as a function of the evolution time in the crystals.} \label{fig:4}
\end{figure}

Now we present experimental tests on the assumption of stationarity\cite{clive1}. We first test the time translation invariance of the system evolution. The input photon's polarization is rotated to $H+V$ ($H-V$) by the HWP. To detect the atomic state in $D$ ($A$) at different readout times $t$, the probe photon state is set as $H+e^{i\varphi_0}V$ ($H-e^{i\varphi_0}V$) with $\varphi_0=-2\pi\delta t$. This photon state is directly mapped to the initial state of atomic excitation. The subsequent system evolution is recorded as the function of the time difference $\tau$. Fig. 3(a) shows the conditional probability
$Q_{ij}(t,t+\tau)$ to find a system in state $j$ at time $t+\tau$, if it is in state $i$ at time $t$. It can be seen that $Q_{ij}(t,t+\tau)$ only depends on time difference $\tau$ and shows little dependence on the evolution time $t$ within experimental errors, thus, the conditional probabilities $Q_{ij}$ are time-translational invariant.

The test on stationarity is completed by assessing the absence of memory effects in our system's evolution.  To do so, we evaluate a measure that is able to detect deviations from Markovian behavior \cite{markov}, as provided by the trace distance $D(\rho_1,\rho_2)=(1/2)tr|\rho_1-\rho_2|$ between any two states $\rho_1$ and $\rho_2$ \cite{breuer}. Markovian processes, which are generally characterized by the absence of information backflow \cite{marcelo}, tend to continuously reduce the distinguishability of physical states and the trace distance should monotonically decrease during the time evolution. To determine the behavior of the trace distance in our system, the two initial states are chosen as $|D\rangle$ and $|A\rangle$, which guarantees the optimality of the measure \cite{markovian}, and makes $D(\rho_1(t),\rho_2(t))=1$ at $t=0$. The atomic states are read-out after different evolution times $t$. Single qubit quantum state tomography \cite{tomography} is carried out to determine the polarization states of the retrieved signal photons. The trace distance $D(\rho_1(t),\rho_2(t))$ is calculated from the reconstructed density matrix of the two states \cite{markovian}. The results are shown in Fig. 3(b), illustrating that the trace distance monotonically decreases in our experiment. This completes the test to verify the assumption of stationarity.

\begin{figure}[tbph]
\begin{center}
\includegraphics [width= 1 \columnwidth]{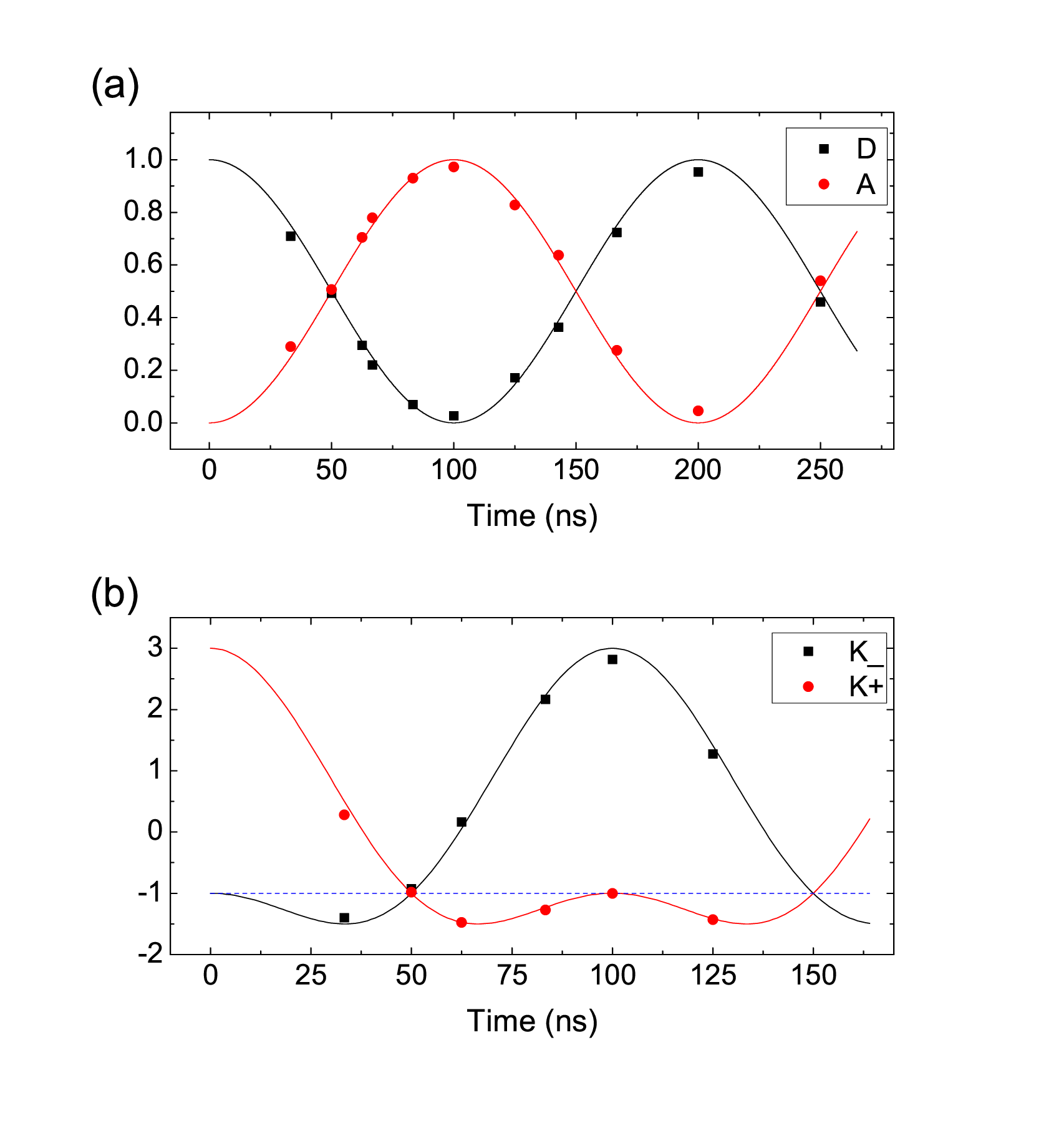}
\end{center}
\caption{(Color online) (a). Time evolution of the probabilities to find the system in atomic state $|D\rangle$ (black squares) and $|A\rangle$ (red dots) for an AFC excitation initially prepared in state $|D\rangle$ and with $\delta = 5$ MHz. (b). The envelope evolution of $\mathbf{K_-}$ (black squares) and $\mathbf{K_+}$ (red dots).  The solid lines are the ideal quantum mechanical predictions and the blue dashed line represents the classical bound.} \label{fig:5}
\end{figure}

We proceed now to demonstrate the violation of the LGtI Eq.(\ref{tbi1}). Fig. 4(a) shows the time evolution of the $|\Psi(t)\rangle$ with an initial state specified by $|D\rangle$. With the AFC prepared with $\delta$ of 5 MHz, the possibility to find the state $|D\rangle$ ($|A\rangle$) oscillates with a periodicity of 200 ns. In agreement with the results in Fig. 3(b), the selected time scale corresponds to a low noise domain with almost complete population inversion. Based on the measured results in Fig. 4(a), the evolution of $\mathbf{K_-}$ and $\mathbf{K_+}$ is shown in Fig. 4(b). $\mathbf{K_+}$ reaches $-1.48\pm0.07$ at 62.5 ns, which violates the classical limit of -1 by 6.9 standard deviations. We also measure the envelope evolution of $\mathbf{K_-}$ with a different evolution speed ($\delta$ = 2 MHz). $\mathbf{K_-}$ reaches $-1.41\pm0.08$ at 83.3 ns.

As a result, our experiment shows clear deviations from the classical bound $K_{\mp}^{min}=-1$. Given that we have independently verified the stationarity of the evolution, the violation of the LGtI Eq.(\ref{tbi1}) is attributed to the coherent nature of the process. In this sense, the presented results address the reverse situation to the one analyzed in Ref. \cite{mauro}, where LGtI are used as non-Markovianity witnesses for the evolution of quantum coherent microscopic systems.

Despite the prepared quantum states involve delocalized excitation over a millimeter scale, attending to the original formulation \cite{macro,lgi,lgi3}, the considered states are of low {\em disconnectivity} ($D=1$) \cite{macro} and can only be treated as a microscopic excitation in macroscopic objects. To substantially improve the macroscopicity, it will be a good choice to combine the atomic memory with the multiphoton entangled source \cite{ghz,entanglement,macro1,macro2}. Given that a single photon entangled with $\sim10^8$ photons has been demonstrated in Ref. \cite{macro2}, it should be possible to prepare superposition states of a macroscopic number of atomic excitations with $D \gg 1$. For future tests on extended LGtI, several current experimental limitations should be considered. The low heralding efficiency of the photon source ($\sim$10\%), the low storage efficiency (several percents) and the nonunit detection efficiency ($\sim$35\%) introduced postselection in the present experiments. With improved mode matching \cite{heralding} and highly efficient single-photon detectors \cite{spd}, a photon source with heralding efficiency greater than 85\% can be expected. A storage efficiency of greater than 70\% has been achieved with rare-earth ion doped crystals with controlled-reversible-inhomogeneous-broadening protocol \cite{eff10}. We expect that the postselection in the current experiment can be avoided in future experiments through the combination of these techniques.

In summary, our experiment demonstrates that the dynamics of a collective excitation that is distributed across two macroscopically separated crystals can be well described by quantum mechanics, while classical realistic theories would need to resort to memory effects to reproduce the observations \cite{montina}. Our controlled independent tests on Markovianity exclude memory effects (at least from the time evolution) in our setup and, thus, make that possibility far less plausible. We note that our experiments cannot falsify Bohmian mechanics \cite{weak2} because this formalism allows for arbitrary superposition states for the optical polarization \cite{bohm}, just like orthodox quantum mechanics. These results provide a general procedure to benchmark ``{\em quantumness}" when temporal correlations can be independently assessed and confirm the persistence of quantum coherence effects in systems of increasing complexity. The polarization dependent AFC technique, which is able to drive the dynamical evolution of collective atomic excitations, may find further applications in both quantum and classical information processing.

{\bf  Acknowledgments}
We are most grateful to A. J. Leggett for his comments on the very preliminary version of this manuscript and to C. Brukner and M. Paternostro for their feedback on a revised version. We also thank D. D\"{u}rr for his kind explanations of the view point underlying a Bohmian approach. This work was supported by the National Basic Research Program (Grant No. 2011CB921200), the Strategic Priority Research Program (B) of the Chinese Academy of Sciences (Grant No. XDB01030300), National Natural Science Foundation of China (Grant Nos. 11274289, 11325419, 61327901, 11504362), the Fundamental Research Funds for the Central Universities (Grant Nos. wk2030380004 and wk2470000011), and the EU STREP projects PAPETS and QUCHIP.

\section*{Supplementary Material}

\textbf{Sample properties.}
The initial absorption depth of the sample is $d\simeq8.0$. With optimized spectral tailoring, we can prepare atomic frequency comb (AFC) with extremely small background absorption $d_0\leq0.05$. The Zeeman population relaxation lifetime is measured by spectral-hole burning, $T_Z=42.9$ ms in the current crystals, which is 7 times longer than that of 10 ppm doped crystals \cite{SHB}. With two-pulse photon echoes, we measured an optical coherence time $T_2=28.4$ us (homogeneous linewidth $\Gamma_h=11$ kHz). The storage efficiency of the quantum memories for bandwidth matched laser pulses at different evolution time is shown in Fig. S1. Because of the better coherent properties and the more efficient optical pumping achieved in the current sample, the storage efficiency is significantly greater than that obtained in 10 ppm doped crystals \cite{polarization}. Note that the crystal only strongly interact with $H$-polarized photons. So the $H$- and $V$- polarized photons with wavelength of approximately 880 nm can be independently processed by the first and second crystal, respectively. Here, the $H$ direction is defined by the crystal's c-axis.

\textbf{Details of the Setup.}
The master laser is a frequency stabilized Ti:Sapphire laser (MBR-110, Coherent) with frequency $\nu_0$ of 340.69656 THz. The frequency of the master laser is stabilized to a low-drift Fabry-P\'{e}rot Interferometer (FPI). The FPI is placed in vacuum housing with temperature controlled to a stability of 5 mK. The linewidth of the master laser is below 50 kHz.

Part of the laser output is fed into a cavity enhanced frequency doubler to generate blue light. The blue light is controlled with an -200-MHz AOM1. The photon pairs are filtered with the optical gratings and the etalon, combined with single-mode fiber (SMF) collection. The signal photons are collected with SMF and ready for use. Another part of the master laser output is used for AFC preparation. The AOM2 with center frequency of -250 MHz is used in double-pass configuration to down shift the laser frequency. The AOM3 and AOM4 have a center frequency of 200 MHz and are also used in double-pass configuration. Thus the center frequency of the AFC preparation matches with the photon source at 340.69646 THz. The pulse sequence for AFC preparation is shown in Fig. 1(b) in the main text, in which the frequency of pump light is swept over 100 MHz in 500 $\mu$s cycle and each frequency point has been assigned specific amplitude to give a comb structure. The resolution of the spectral tailoring is approximately 500 kHz. The AFC storage time is determined by $1/\Delta$ so only specific storage times can be probed. The tests on stationarity in Fig. 3(a) require measurements at the time t and the time t + 33.3/66.7/100 ns, so the choices for storage times are severely limited and little measurement results are shown for t$>$100ns. Nevertheless, the tested timescale already covered the time range for violations of LGtI.
\begin{figure}[tb]
\centering
\includegraphics[width=0.5\textwidth]{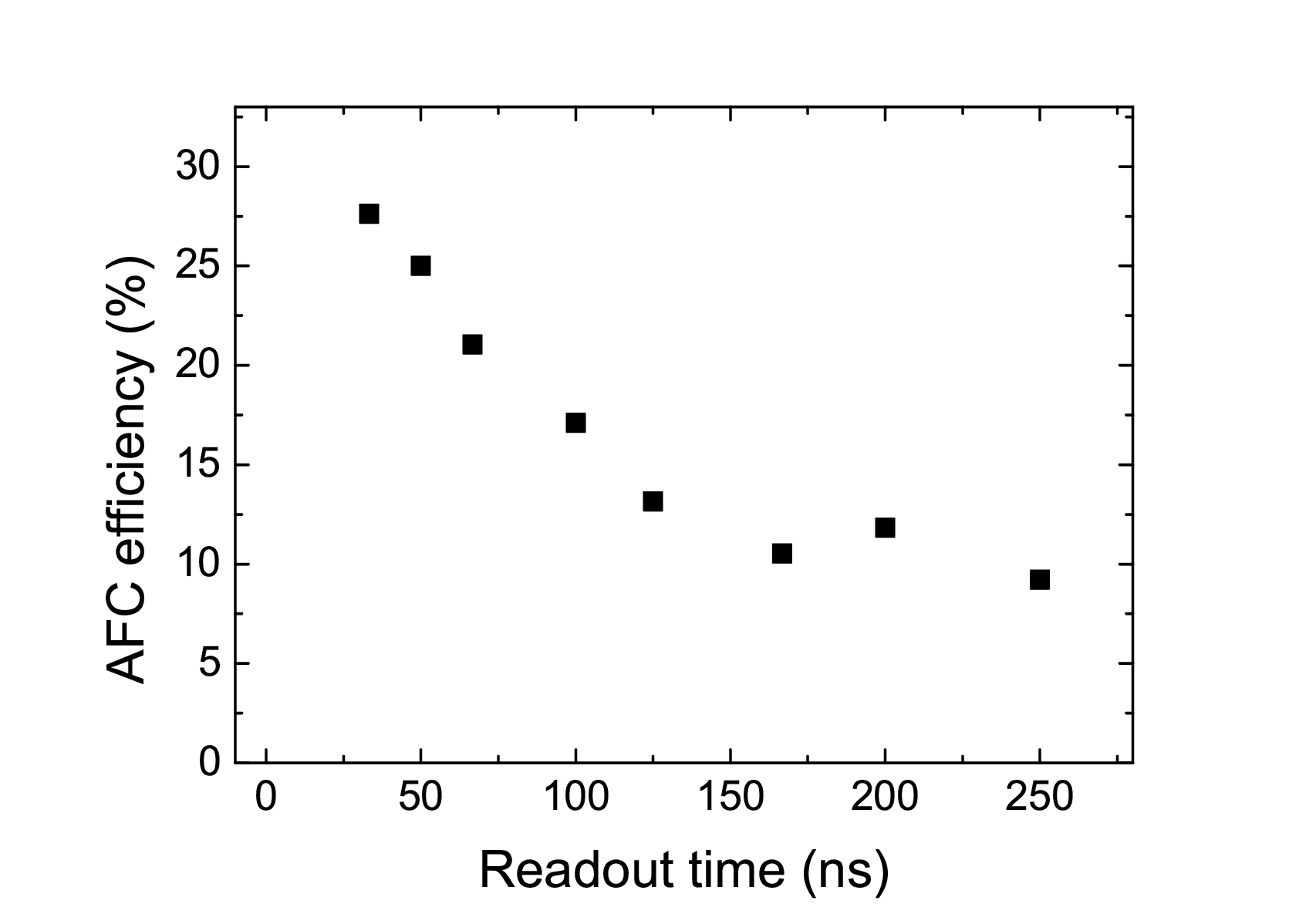}
\\\begin{flushleft}
Fig. S1: The readout efficiency for bandwidth-matched laser pulses at different times. The AFC efficiency is averaged over the measurement phase.
\end{flushleft}
\end{figure}

\begin{figure*}[tb]
\centering
\includegraphics[width=0.7\textwidth]{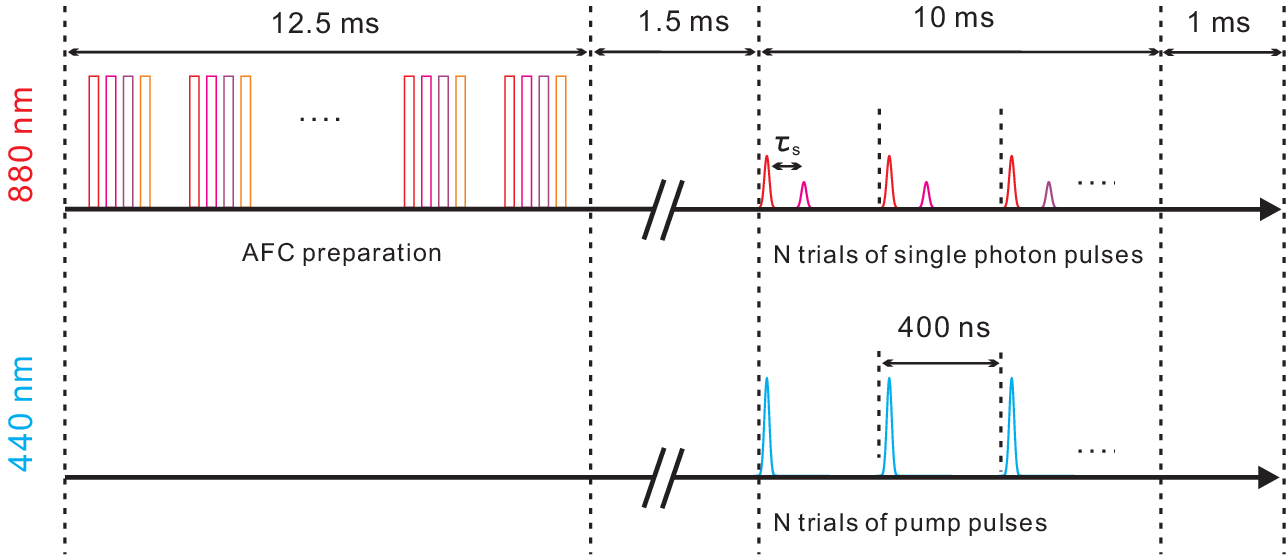}
\\\begin{flushleft}
Fig. S2: The timing sequence used for the experiment. The complete preparation and measurement cycles are repeated at a frequency of 40 Hz.
\end{flushleft}
\end{figure*}
The AFC preparation light and signal photons are combined with 10:90 beam splitter and collected with SMF. The light focus to a diameter of 100 $\mu m$ with a lens (f=250 mm). The sample is placed in a cryostat at a temperature of 1.5 K and with a superconducting magnetic field of 0.3 T in the $H$ direction. The two parallel Nd$^{3+}$:YVO$_4$ crystals' c-axes are placed in the $H$ direction. Because each crystal only strongly absorbs $H$-polarized light, the $H$-polarized components of input photons are stored in the first crystal, and the $V$-polarized components are stored in the second crystal after polarization rotation by the HWP. The single photons are detected with an efficiency of approximately 35\% with 50-Hz dark counts.

The preparation and measurement timing is controlled by three arbitrary function generators (Tektronix, AFG3252). As shown in Fig. S2, the AFC preparation takes 12.5 ms. To avoid the fluorescence noise caused by the classical pump light, The measurement phase starts 1.5 ms after the preparation completes. $N$ trials of single photon pulses are stored in the sample in the 10-ms measurement phase. The complete preparation and measurement cycles are repeated at a frequency of 40 Hz.  The chopper before SPD is opened in the measurement phases to enable detection of single photons.

During the measurement cycle, $N=25000$ trials of 440 nm laser pulses are sent into the PPKTP crystal with a periodicity of 400 ns. The generated single photon pulses are stored in the sample and retrieved after a programmable time $\tau_s$. The pulsed operation of the photon source further improves the signal-to-noise ratio as a result of the temporal separation of the retrieved photons from the transmitted noise. Due to the mismatched bandwidth of the signal photons and the quantum memories, most of the input photons are transmitted directly through the sample. The extinction ratio for orthogonal polarizations is approximately 1000:1 in our setup, so a small transmission peak still can be seen at $t=0$ in Fig. S4(a). For the measurements presented in Fig. 3(b), to probe the quantum states for longer storage time, $N=10000$ trials of single-photon pulses are sent into the sample with a periodicity of 1000 ns.

\begin{figure}[tb]
\centering
\includegraphics[width=0.4\textwidth]{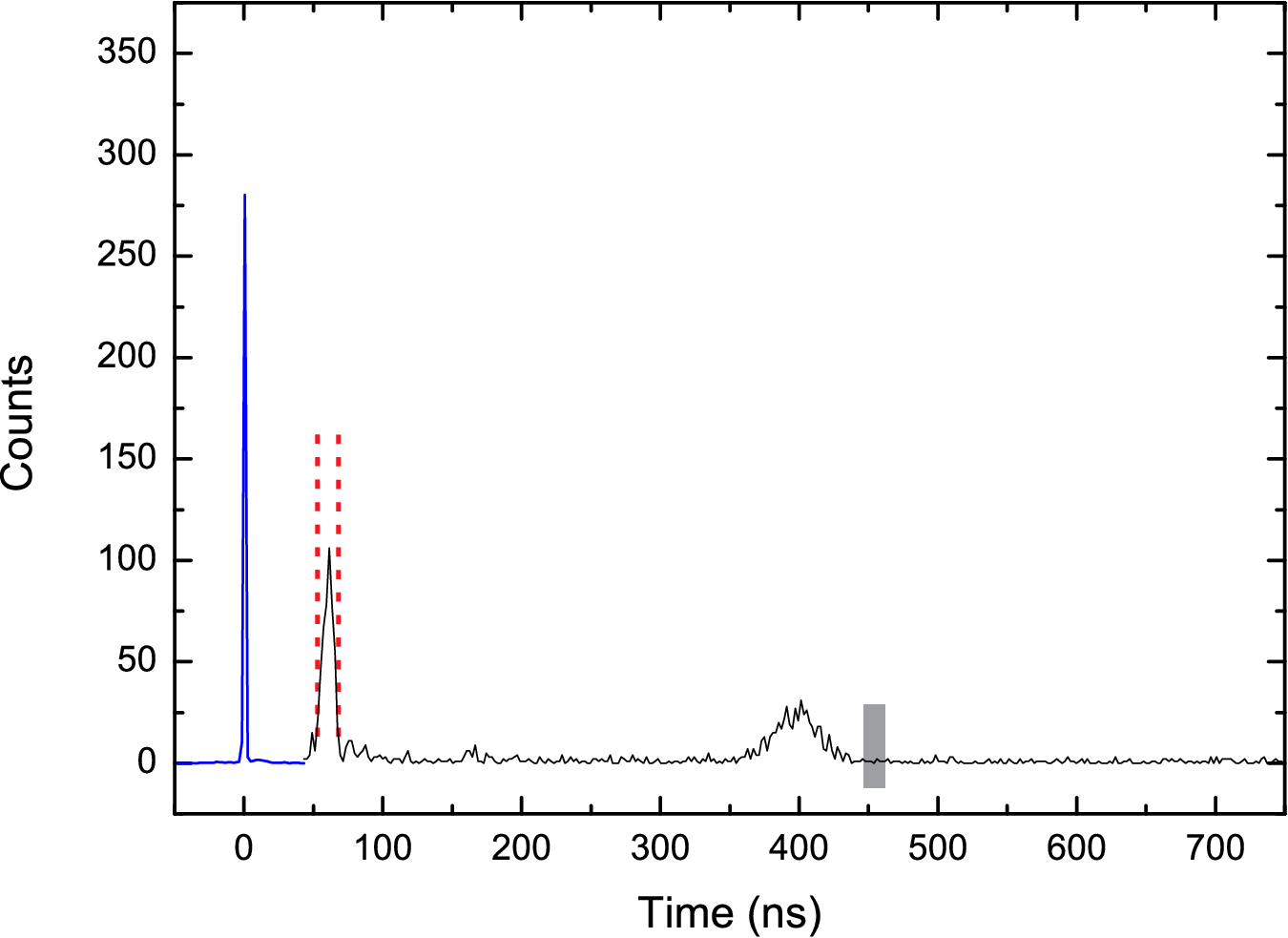}
\\\begin{flushleft}
Fig. S3: Measurement on $g^{(2)}_{s i}$ after a storage time of 50 ns. The blue parts shows the transmitted components which have been intentionally decreased by 50 times for visual effect. The red dashed lines and shaded area define the 10-ns detection windows for the coincidences and noise used to compute cross-correlations of the retrieved signal, respectively. The time-bin size is 2 ns and the integration time is 10 minutes.
\end{flushleft}
\end{figure}

\begin{figure*}[tb]
\centering
\includegraphics[width=1.0\textwidth]{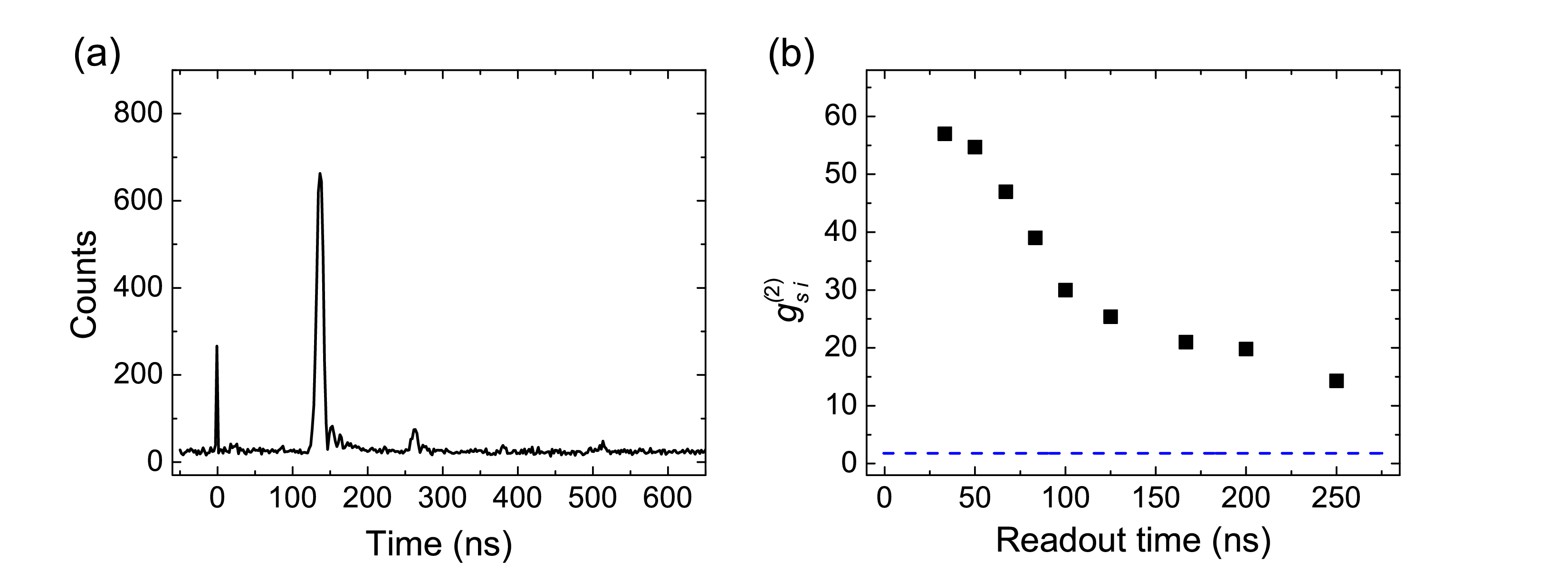}
\\\begin{flushleft}
Fig. S4: {\bf a}, Example of storage of the $H+V$ polarized photons in the polarization dependent AFC with readout polarization of $H-V$. Most parts of the transmitted parts are blocked by the PBS. A strong echo is emitted at 125 ns with rotated polarization states. The time-bin size is 2 ns and the integration time is 4 hours. {\bf b}, Cross-correlations $g^{(2)}_{s i}$ as a function of storage time. The blue dashed lines corresponds to the classical limit $g^{(2)}_{s i}=2$ for two-mode squeezed states.
\end{flushleft}
\end{figure*}
\textbf{Details of the photon source.}
The phase matching temperature to generate degenerate photon pairs is 44.0 $^{\circ}$C.
The power of the 440 nm pump light is 20 mW for all the measurements. With continuous pump, the coincidence rate of the unfiltered photon pairs is 1.4 MHz. The heralding efficiency is 0.20. The diffraction efficiency of the gratings (1200 lines per millimeter) is 0.88. The transmission efficiency of the etalon is 0.95. The etalon has a bandwidth of 700 MHz and a free spectral range of 50 GHz. The temperature drift of the etalon is smaller than 10 mK. After spectral filtering, the coincidence rate of the photon pairs is 6.1 kHz and the heralding efficiency is 0.10. The photon source is switched into pulsed operations for interface with the AFC. The temporal width of blue pump light is 32 ns. The transmission efficiency from the photon source to the final SPD for the signal photons is approximately 0.23.

\textbf{Heralded storage of single photons.}
A figure of merit for the nonclassical nature of the photon correlations is the normalized cross-correlation function $g^{(2)}_{s i}=p_{s i}/p_sp_i$, where $p_s$ ($p_i$) is the probability to detect a signal (idler) photon and $p_{s i}$ is the probability to detect a coincidence in a specified time window. In practice, $p_{s i}$ and $p_sp_i$ are determined by the number of coincidences in the time window centered on and away from the coincidence peak, respectively. Assuming that the second-order auto-correlations of signal and idler $g^{(2)}_{x}$ (where x=`s' for signal or `i' for idler) satisfying $1\leq g^{(2)}_{x} \leq2$, then the non-classicality is proved by measuring $g^{(2)}_{s i}$ greater than 2 \cite{entangle,ref1,g2}.

We now verify that the non-classical correlations between signal and idler photons are preserved during the storage and retrieval process. Fig. S4(a) shows an example of storage the $H+V$-polarized single photons and readout with $H-V$ polarizations after 125 ns delay. The frequency detuning of two AFC is set as 5 MHz. A strong echo is emitted after 125 ns. Due to the 100-MHz bandwidth of the AFC, the retrieved photons has a temporal width of approximately 10 ns. The cross-correlation function $g^{(2)}_{s i}$ is measured in the the $H$ or $V$ polarization basis to avoid polarization rotation caused by the detuned AFC.

An example of measurement on $g^{(2)}_{s i}$ is shown in Fig. S3. $V$-polarized single-photon pulses are mapped into the sample and retrieved 50 ns later. The noise at 400 ns is from the next pump pulse of SPDC. To determine the $g^{(2)}_{s i}$ of the transmitted pulses, the 2-ns coincidence window is placed at $t=0$ to measure the signal and the detection window for related noise is placed at $t=400$ ns. Because of the temporal broadening of the retrieved signal pulses, the coincidence window for retrieved signal has the width of 10 ns. The detection window for related noise is also placed at 400 ns away from the retrieved signal. We measured $g^{(2)}_{s i}=54.7$ after a storage time of 50 ns.

The measured results of $g^{(2)}_{s i}$  are shown in Fig. S4(b). Since the values of $g^{(2)}_{s i}$ at all times are significantly greater than the classical bound of 2, any classical description of light is eliminated. We measured $g^{(2)}_{s i} \approx 452$ between the transmitted signal photon and the idler photon which allows to upper bound the autocorrelation of the heralded signal photon to $g^{(2)}_{i|ss}\leq0.0088$ \cite{entangle}. This indicates that the photon source itself is close to ideal single photons ($g^{(2)}_{i|ss}=0$) and ensures that there can hardly be excitations of more than one in the crystals. We note that, compared with the unfiltered source, $g^{(2)}_{s i}$ rises significantly after spectral filtering. The $g^{(2)}_{s i}$ drops after storage because of low recall efficiency, the temporal broadening and the rising contributions of dark counts.

\end{document}